\documentstyle[prl,aps,psfig]{revtex}

\begin{document}
\draft

\title{Spin separation in digital ferromagnetic heterostructures}

\twocolumn[\hsize\textwidth\columnwidth\hsize\csname@twocolumnfalse\endcsname

\author{J. Fern\'andez-Rossier$^{*}$ and L. J. Sham}
\address{Department of Physics, University of California San Diego, \\
9500 Gilman Drive, La Jolla, CA 92093}

\date{\today}
\maketitle

\begin{abstract}

In a study of the ferromagnetic phase of a multilayer digital ferromagnetic
semiconductor in the
mean-field and effective-mass approximations, we find the exchange
interaction to have the
dominant energy scale of the problem, effectively controlling the spatial
distribution of the  carrier
spins in the digital ferromagnetic heterostructures.  In the ferromagnetic
phase, the majority
and minority carriers tend to be in different regions of
the space (spin separation). Hence, the charge distribution of carriers also
changes noticeably from the ferromagnetic to the paramagnetic
phase. An example of a design to exploit these phenomena is given.

PACS number(s):  75.70.Cn, 75.50.Pp, 75.10.-b
\end{abstract}
]
%\begin{multicols}{2}
\narrowtext

The research in proactive use of the  spins of the carriers to add a new
dimension to electronics starts a new area known as
spintronics.\cite{Spintronics}  The recent discoveries \cite{Ohno96,GaNMn} of
ferromagnetism with high Curie temperatures in a number of conventional
semiconductors doped with magnetic impurities hold promise for the
implementation of spintronics in semiconductors.

Inhomogeneously doped semiconductors, such as the p-n junction,  play a crucial
role in conventional electronic devices. Their properties  depend on the
distribution of the itinerant carriers, which is governed by the Coulomb
interaction of the carriers with the impurities and with other carriers.   In
this paper we study the charge and spin  distributions of the itinerant
carriers in semiconductors delta-doped with magnetic impurities, such as
GaMnAs.\cite{Kawakami,Chen} Our theory is within the mean-field, effective-mass
and virtual-crystal approximations. In addition to the electrostatic forces,
the itinerant carriers have an exchange interaction with the magnetic
impurities. Our calculation shows that the magnetization of the high Mn
concentration in the delta layers in the ferromagnetic phase gives rise to a
spin-dependent potential experienced by the carriers  comparable in order of
magnitude to the charge potential of the delta layer. The effect of the
spin-dependent potential on the inhomogeneous spin distribution of the 
carriers was noted by Loureriro da Silva  {\em et al.} \cite{cunha} in multilayered GaMnAs
with 5\% magnetic impurities.  By contrast, the delta doping with a nominal
concentration per atomic plane of 25\%  to 50\% Mn atoms gives rise to a
qualitative different phenomenon of spin separation. This creates the
possibility of the magnetic control of the itinerant carriers by manipulating
the magnetization of the Mn ions. In particular, the spin-dependent potential
may be used to influence the spin dependence of the distribution of itinerant
carriers. The majority spin carriers accumulate in the region of the Mn layers
whereas minority carriers are repelled from the Mn region.   In the
paramagnetic phase, the spin potential averages to zero and the magnetic
influence  disappears.  This prediction of large changes in the distribution of
carriers between the paramagnetic and the ferromagnetic phases may be a cause
of the anomalous Hall effect in the ferromagnetic phase observed in the
delta-doped systems.\cite{beth}   To show the potential of the spin separation for device
applications, we give an example of how heterostructures may be designed to
engender spin separation.

This paper is organized as follows. A brief review of the relevant theoretical
formalism precedes the application to establish the general principle of
magnetic control and to show how it  works in two specific systems.  The first
is  the experimental system of multiple digital layers of references
\onlinecite{Kawakami,Chen}. We find the phenomenon of spin separation in the
ferromagnetic phase.  Specifically, for appropriate carrier densities and
interlayer distances, the minority carriers are located mainly in the
interlayer spacer, with a very small overlap with the magnetic atoms whereas
the majority carriers are mainly located in the layers of magnetic atoms.
Second, we design a double GaAs quantum well with AlGaAs barriers and a delta
layer of Mn in the middle of one well and a delta layer of an acceptor (Be) in
the middle of the other.  When the GaMnAs well is changed from the paramagnetic
phase to the ferromagnetic phase, a net transfer of charge and spin from the Be
well to the Mn well is produced, because of the appearance of the dominant
magnetic interaction. As a result, there is a potential drop across the
structure as well as a spin polarization in the Be well.

%{\em The model}.
 Our calculations are based on the mean-field  model,
\cite{Mac99,Mac00,nosoqw,Dietl00,us,cunha} with two types of spins: localized d
electrons with magnetic moments of $5/2$ Bohr magnetons and itinerant carriers
(holes because of the Mn and Be doping). There is a Heisenberg spin exchange
between the  itinerant carrier and the Mn d electron, which  is responsible for
the ferromagnetism. The hole energy bands are based on the effective mass
approximation with the kinetic energy given by the bands of the host GaAs.  The
effective potential induced by the dopants is the only change for the carrier
subbands,  as in the virtual crystal approximation.  The Coulomb interaction
between the holes is taken into account in the Hartree approximation.  The
in-plane energy dispersion of the holes subbands is described  by a parabolic
model which neglects the spin-orbit interaction. The omission of the spin-orbit
effect, which has also been used in references
\onlinecite{cunha,Mac99,Mac00,nosoqw}, greatly simplifies the numerical
computation.  On the other hand, it leads to an overestimate of the
polarization of the Mn spins induced by the interaction with the carriers. We
correct for this by using a smaller value of the coupling constant $J$ of the
Heisenberg exchange which reproduces the Curie temperature obtained in our
previous calculation including the spin-orbit interaction.\cite{us} Because the
charge and spin distributions are averaged properties of the system, we argue
that the qualitative aspects of the effects discussed in this paper would
remain if the spin-orbit interaction is included.

Both model systems studied here are taken to be translationally invariant in
the $xy$-plane normal to the growth axis. Thus, the Mn concentration $c_M(z)$ 
varies only along the $z$ axis. We assume that, in addition to the magnetic
impurities which act as  acceptors in GaMnAs, there is a distribution of
donors, $c_c(z)$, which partially compensate the Mn acceptors, so that the
total density of holes is smaller than the density of Mn, as is observed.
\cite{Ohno96} For simplicity, we assume that the spatial distributions of the
donor and acceptor impurities are the same except for a multiplicative
constant. From the charge neutrality condition, $P+C_c=C_M$, where $P$, $C_c$
and $C_M$ are the average densities of holes, of the compensating impurities,
and of the Mn impurities, respectively.

In the mean-field approximation, the effect of the magnetic impurities on the itinerant carriers is
given by a spin-dependent potential:
\begin{equation} V_
{\sigma} = \frac{J\;\sigma}{2} c_M(z) \langle M(z) \rangle
\label{sppot}
\end{equation} where $\sigma=\pm 1$ denotes the spin directions, $J$ is the
exchange coupling constant  between the itinerant carrier and the Mn spin, and
$\langle M(z) \rangle$ stands for the local average magnetization of  Mn. The
Mn polarization is produced by the molecular field created by the spin
polarization of the itinerant carriers as well as the external  magnetic field.
At  temperature $T$ and zero external magnetic field, the Mn magnetization is
given by the usual Brillouin function:
\begin{equation} M(z)=S\; B_S\left( \frac{J}{k_BT}
\frac{p_{+}(z)-p_{-}(z)}{2}
 \right)
\end{equation} where $p_{\sigma}(z)$ is the spin-dependent density of the
itinerant carriers.

The effective Schr\"odinger equation for the holes has a self-consistent
potential $V_e(z)+V_{\sigma}(z)$, where $V_e(z)$ is the electrostatic potential
seen by the hole given by,
\begin{equation}
\frac{d^2}{dz^2}V_e(z)=\frac{4\pi\;e^2}{\epsilon}[c_c(z)-c_M(z)+p_+(z)+p_-(z)]
\end{equation}
 The hole eigenenergy is the sum of the quantized energy $E_{n,\sigma}$ from
the motion in the $z$ direction and the plane wave energy
$\epsilon_{\vec{k}_\parallel}=\frac{\hbar^2k^2_\parallel}{2m_\parallel}$ from
the inplane motion. The hole wave function  is the product of the bound state
envelope function in $z$ times the plane wave normal to the $z$ axis.  The
spin-dependent hole density is given by:
\begin{equation} p_{\sigma}(z) = \sum_n \int \frac{d^2
\vec{k}_\parallel}{(2 \pi)^2}
\frac{|\phi_{n,\sigma}(z)|^2}{e^{(E_{n,\sigma}+
\epsilon_{\vec{k}_\parallel} -\mu)/k_BT} +1} ,
\end{equation} where $\mu$ is the chemical potential.

It is illuminating to estimate  the strength of the spin dependent potential
(\ref{sppot}) for a single magnetic digital layer.  Throughout the paper we
take the hole effective mass tensor to be $m_\parallel=0.11 $ and $m_z=0.37$
times the free electron mass and model the distribution of the Mn in a digital
layer as Gaussian:
\begin{equation} c_M(z)=\frac{ C_M}{\sqrt{\pi}
\Delta}Exp[-(z/\Delta)^2]
\label{gauss}
\end{equation} 
In a digital layer of Ga$_{0.5}$Mn$_{0.5}$As the total concentration of Mn is
$C_M=3.13\times 10^{14} \mbox{~cm}^{-2} = 3.13 \mbox{~nm}^{-2}$. Transmission
electron microscopy experiments \cite{Kawakami} reveal that the Mn is spread
over 2 or 3 atomic planes. This corresponds roughly to the Gaussian half-width
of $\Delta=0.5$~nm. The exchange integral of $J=150$ mev-nm$^3$ used in our
previous work\cite{us} to calculate the Curie temperature is reduced to
$J=100$ mev-nm$^3$ here in order to compensate the  absence of spin-orbit in
the present calculation.  The maximum of the spin-dependent potential in
Eq.~(\ref{sppot}) for saturated magnetization is  441.5~meV in comparison with
the charge well depth of about 177~meV.  The peak to valley splitting is twice the maximum value. In a diluted
magnetic semiconductor quantum well, the magnetic potential is one order of
magnitude smaller than in a digital layer. Hence, the magnetic control is more
practical in digital layers  than in quantum wells.

\begin{figure}
\centerline{\psfig{figure=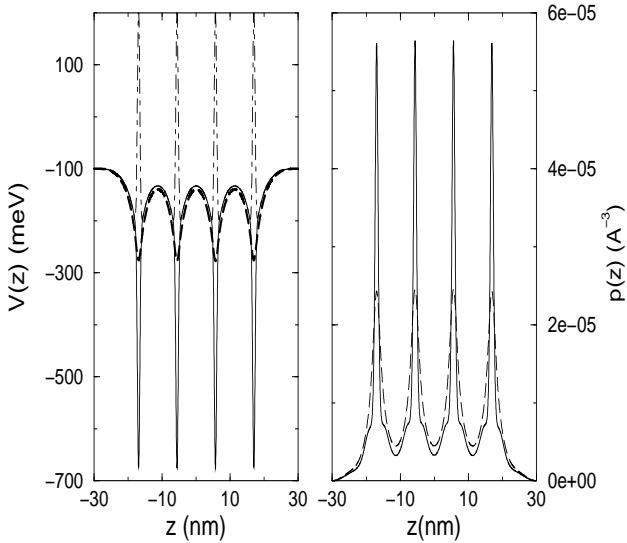,height=3.0in,width=3.3in}}
\caption{ Left panel: spin-dependent potentials for majority (solid line)
and minority (thin
dot-dashed line)  at T=5K (the ferromagnetic phase) and  total potential at
T=40 K (the
paramagnetic phase) (thick dashed line). Right panel: total carrier
density profiles for T=5K (solid
line, the ferromagnetic case) for T=40 K (long-dashed line, the
paramagnetic case).  }
\label{fig1}
\end{figure}

The magnetic potential is attractive for the majority carriers and repulsive
for the minority carriers. As the temperature is increased, the magnetization
decreases and so  does the magnetic potential, vanishing above the Curie
temperature. Accordingly, the itinerant carrier density profiles change
significantly as the temperature varies. Both the potentials and the density
profiles are shown in Fig.~\ref{fig1} for a system of $N=4$ digital layers, 
with an interlayer separation $D=40$~ML (GaAs monolayers), or 11.3~nm,  and a
density of  holes per layer $p=1.3\cdot10^{13}$~cm$^{-2}$.  Similar results are
obtained for systems between 1 and 10 layers. In the left  panel of
Fig.~\ref{fig1} we present the potential profiles at $T=$40 K (thick dashed
line), when the system is paramagnetic. In the same panel we show the total
potential for majority (attractive) and minority (repulsive) carriers for spin
dependent potentials at $T=$5 K  when the system is ferromagnetic. The large
magnitude of the spin-dependent part of the potential is apparent in the
figure. The effect on the density profile of the itinerant carriers is shown in
the right panel. In the ferromagnetic case (thin solid line) the carriers tend
to pile up in the magnetic layers. In the paramagnetic case (long dashed line)
the carrier distribution  is more spread out, with more carriers in the spacer
layers relative to the ferromagnetic case.

\begin{figure}
\centerline{\psfig{figure=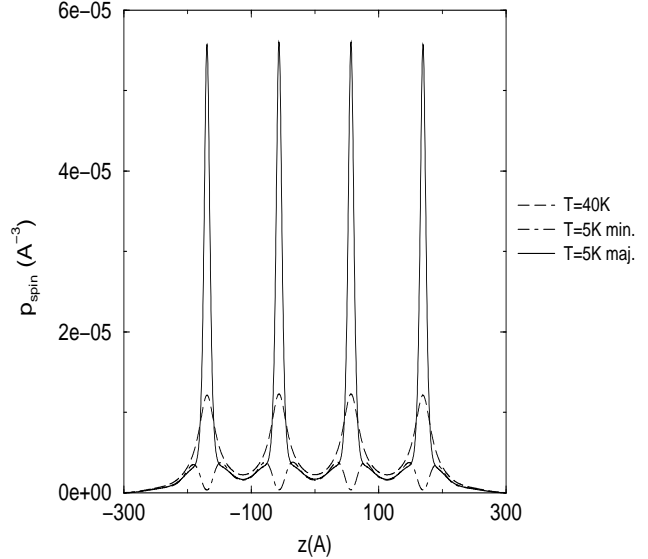,height=3.0in,width=3.3in}}
\caption{  Spin dependent hole density profiles for T=5K majority carriers
(solid), minority carriers
(long dashed) and for T=40 K (long-dashed). }
\label{fig2}
\end{figure}

 Figure~\ref{fig2}  shows the resultant spin-dependent density profiles for
two temperatures.   In
the ferromagnetic case, the distributions of  majority and minority
carriers are distinct. The majority
carriers are localized in the magnetic layers whereas the minority carriers
are almost totally
expelled from it. This ``spin separation'' phenomenon is, of course, absent
in the paramagnetic
case.
  We have checked that as the interlayer distance decreases or the hole density
 increases the spin separation diminishes. Spin fluctuations of Mn beyond the
 mean-field approximation are a major source of spin-flip scattering. Hence, we
 expect the spin-flip scattering to be reduced when the spin separation is
 larger. We have also found a correlation between the absence (or presence)
of the observed
anomalous Hall effect in transport experiments \cite{beth} and the presence
(or absence) of the
spin separation in our calculations of the same sample configurations. The
anticorrelation may
provide a clue to construct in future a theory of the anomalous Hall effect
based on the absence of
spin separation.

Results of  figure (\ref{fig1}) show how the exchange interaction can
overcome the electrostatic
interactions in the case of a digital layer. We propose here a
heterostructure in which this effect is
made more apparent and can  be measured.  It is a double quantum well of
GaAs with GaAsAl
barriers, delta-doped with one layer of Mn in the middle of the left well
and one layer of the
acceptor Be in the middle of the right well. For illustration but not
necessity, we choose the
density profile of the Be to be identical to that of the Mn ions plus its
compensating charges. We
take the density of holes per digital layer in either well of
$1.3 \times 10^{13} cm^{-2}$. The digital layer of Mn has a total density
of $3.13 \times
10^{13} cm^{-2}$ and a Gaussian distribution as in Eq.~(\ref{gauss}) with
$\Delta=0.5$~nm.

\begin{figure}
\centerline{\psfig{figure=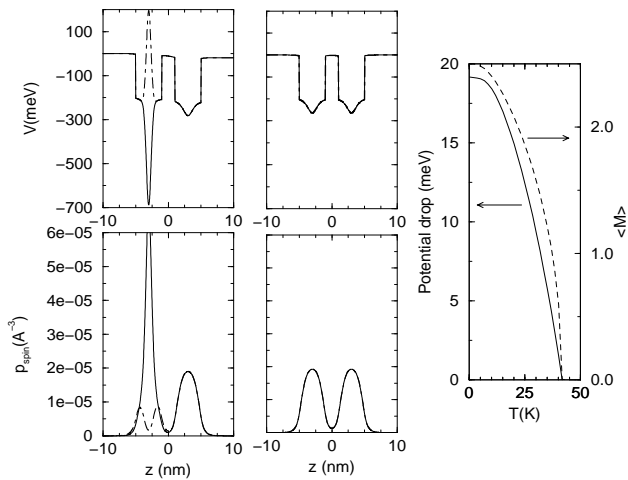,height=2.5in,width=3.3in}}
\caption{ Left upper/lower panels: spin-dependent potentials/density
profiles for majority holes
(solid line) and minority holes (thin dot-dashed line)  at T=5K (the
ferromagnetic phase).  The
middle panels: spin-dependent potentials and density distributions at T=40
K (the paramagnetic
phase) -- same notation as above. Right panel:  potential drop (solid line)
and magnetization
(dashed line) versus temperature. }
\label{fig3}
\end{figure}

Above the Curie temperature the system is totally symmetric (by the
construction of the model) and so is the charge distribution (shown in the
middle column of Fig.~\ref{fig3}).  At low temperatures the majority carriers
with spin antiparallel to the Mn are attracted by the magnetic layer. As a
result, there is a spin-dependent charge transfer from the non-magnetic to the
magnetic well (the left column of Fig.~\ref{fig3}), which generates an internal
electric field and a potential drop. In the rightmost panel we plot both the
potential drop and the Mn average magnetization as a function of the
temperature. The potential drops follows the average magnetization and its
maximum  is close to 20 meV at zero temperature. Larger values can be obtained
for smaller carrier densities, but this decreases the Curie temperature
\cite{us}. The charge imbalance is  $1.2 \times 10^{12} cm^{-2}$ or 4.6$\%$ of
the total hole density.  The charge transfer effect takes also place for other
values of the carrier density, as well as for structures with a different
density of Be and Mn. As a result of the rearrangment of the carriers in the
heterostructure, as the system goes from the paramagnetic to the ferromagnetic
phase, large changes in the in plane and vertical transport properties can be
expected as well.

  Recent Monte Carlo simulations for bulk ferromagnetic semiconductors
\cite{breymc}, in which the the Mn magnetic moment is treated as a classical
vector of fixed norm,  confirm that the carriers are more localized around the
magnetic atoms in the ordered phase than in the paramagnetic phase. This
indicates that the attraction of the carriers to the magnetic atoms is a
cooperative effect  which is accounted for in our mean field calculation. Tight
binding calculations for a superlattice of digital GaAs/GaAsMn \cite{meyer}
give very similar results to those obtained using an envelope function approach
\cite{us}, validating the use of the effective mass approximation in
delta-doped systems.  A density functional calculation  for a superlattice of
digital GaAs/MnAs predicts a half-metallic behavior \cite{ab}. Our results
should be compared with a similar calculation including compensating
impurities, 50\% of Mn and larger interlayer  distances. 

In conclusion, we have studied the spatial distribution of the carrier spins in
digital ferromagnetic heterostructures. Our main results  are: (i) the
distribution of carriers in digital ferromagnetic heterostructures of GaAsMn is
largely controlled by the exchange interaction, which overcomes the
electrostatic interaction. (ii) Large changes in the spatial distribution of
carriers take place when ferromagnetism is switched off (for instance, by
increasing the temperature above the Curie temperature). (iii) In some
situations the minority carriers  are totally separated from the Mn layers and
the majority carriers. This might have consequences in transport properties and
the anomalous Hall effect.

We wish to thank D.D. Awschalom, R. Kawakami, A. Gossard, E. Gwinn, L. Brey and
B. H. Lee for  stimulating discussions. We acknowledge the Spanish Ministry of
Education for a post-doctoral fellowship and support by DARPA/ONR
N0014-99-1-109  and NSF DMR 0099572.

(*) Present Address: University of Texas at Austin, Austin, TX 78712.

%\widetext

%\end{multicols}
\end{document}